\documentclass[final,twocolumn]{IEEEtran}
\usepackage{cite}

\ifCLASSINFOpdf
  \usepackage[pdftex]{graphicx}
\else
  \usepackage[dvipdf]{graphicx}
  \DeclareGraphicsExtensions{.eps}
\fi
\usepackage{amsmath}
\usepackage{fixmath}
\usepackage{amsbsy}
\usepackage{amssymb}
\usepackage{url}

\usepackage{hyphenat}
\usepackage[mediumspace,mediumqspace,squaren]{SIunits}
\usepackage[utf8]{inputenc}

\begin{document}

\title{Automatic Spatial Calibration of Ultra-Low-Field MRI for High-Accuracy Hybrid MEG--MRI}

\author{\IEEEauthorblockN{Antti J.~Mäkinen*, Koos C.~J.~Zevenhoven*, and Risto J.~Ilmoniemi}\thanks{This project has received funding from the International Doctoral Programme in Biomedical Engineering and Medical Physics (iBioMEP), from the Finnish Cultural Foundation, from the Academy of Finland,  and from the European Union's Horizon 2020 research and innovation programme under grant agreement No 686865.}\thanks{* A.~J.~Mäkinen and K.~C.~J.~Zevenhoven contributed equally to this work.} \thanks{All authors are with the Department of Neuroscience and Biomedical Engineering
Aalto University School of Science,
address: P. O. Box 12200, FI-00076 AALTO,
email: antti.makinen@aalto.fi,
phone: +358445713188}\thanks{Copyright (c) 2019 IEEE. Personal use of this material is permitted. However, permission to use this material for any other purposes must be obtained from the IEEE by sending a request to pubs-permissions@ieee.org.}}

\markboth{Mäkinen \MakeLowercase{\textit{et al.}}}{}
%



\maketitle

\begin{abstract}
\noindent
With a hybrid MEG--MRI device that uses the same sensors for both modalities, the co\hyp{}registration of MRI and MEG data can be replaced by an automatic calibration step. Based on the highly accurate signal model of ultra-low-field (ULF) MRI, we introduce a calibration method that eliminates the error sources of traditional co-registration. The signal model includes complex sensitivity profiles of the superconducting pickup coils. In ULF MRI, the profiles are independent of the sample and therefore well-defined. In the most basic form, the spatial information of the profiles, captured in parallel ULF-MR acquisitions, is used to find the exact coordinate transformation required. We assessed our calibration method by simulations assuming a helmet-shaped pick\-up-coil-array geometry. Using a carefully constructed objective function and sufficient approximations, even with low-SNR images, sub-voxel and sub\hyp{}millimeter calibration accuracy was achieved. After the calibration, distortion-free MRI and high spatial accuracy for MEG source localization can be achieved. For an accurate sensor-array geometry, the co-registration and associated errors are eliminated, and the positional error can be reduced to a negligible level.
\end{abstract}

%


\section{Introduction}
\noindent
In magnetoencephalography (MEG), neuronal activity in the brain is estimated from magnetic field measurements with an array of sensors outside the head \cite{Hamalainen1993}. As the overall precision of MEG increases, accurate knowledge of the brain conductivity structure with respect to the magnetic field sensors becomes more and more important for modeling the neuronal fields at the sensors. Independent of this, accurate positional information of the brain anatomy is also essential when setting constraints for reconstructing the neuronal sources \cite{Lin2006}. This information usually comes from magnetic resonance (MR) images of the subject's head. Because the brain structure is imaged in a different device, the MR images and the MEG sensor array have to be aligned, or \textit{co\hyp{}registered}, with respect to each other.

Conventional co\hyp{}registration procedures involve manual steps with several possible error sources \cite{Whalen2008,Bardouille2012}, complicating the MEG workflow and data analysis. Inaccuracies in the co\hyp{}registration distort both the conductor model and the source model of the neuromagnetic problem, causing errors in the localization of brain activity. The total co-registration error can be up to 10 mm and, depending on assumptions in the source modeling, it can significantly deteriorate the inverse estimates \cite{Hillebrand2011}. Improving the co\hyp{}registration accuracy in a systematic fashion \cite{Adjamian2004,Troebinger2014a} has thus far involved external equipment, further complicating the workflow.

\begin{figure*}[tb]
    \centering
    \includegraphics[width=\textwidth]{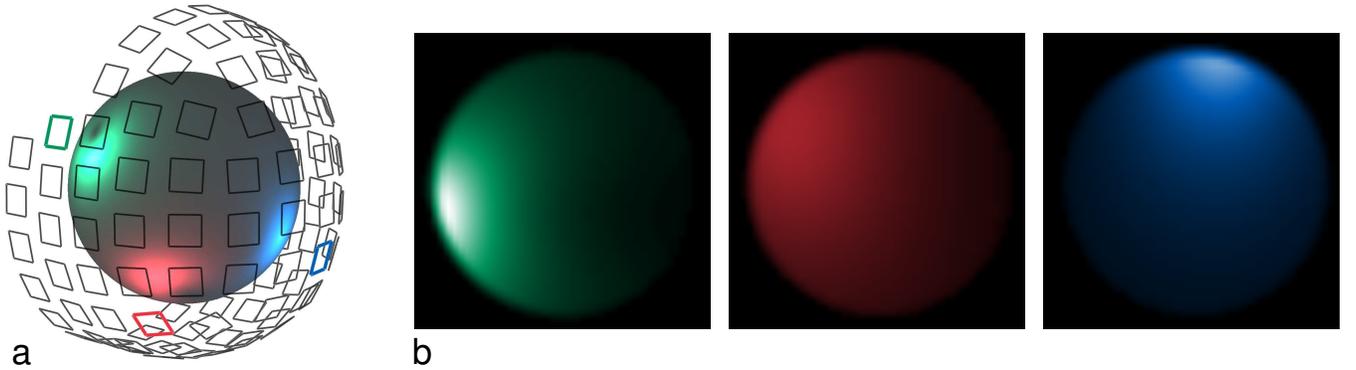}
    \caption{Visualization of calculated sensitivity profiles in the array frame of reference and in MRI for three selected sensors. (a) Selected SQUID magnetometer pickup coils (red, green, and blue) in the helmet-shaped sensor array. Absolute values of sensitivity profiles $|\beta_j(\vec{r}\,)|$ of the three colored pickups are plotted with corresponding colors on the surface of a spherical phantom inside the array. (b)  Illustration of the same slice from three single-coil MR images corresponding the colored pickups in Fig (a). The profiles $|\beta_j(\boldsymbol q _n)|$ can be observed in the phantom volume. Voxels with maximal intensity can be found at the boundary of the phantom closest the corresponding pickup coil.}
    \label{fig:profiles}
\end{figure*}

With a hybrid MEG--MRI device \cite{Vesanen2013}, a co\hyp{}registration-free workflow can be achieved, when both modalities are measured using the same sensor array. Instead of separately registering an MRI with the MEG array for each individual subject, the hybrid device can be calibrated so that the MRI is automatically reconstructed in the same coordinate system as the MEG. 
In this work, we introduce a calibration method that utilizes spatial information in the MR \textit{sensitivity profiles} of the sensor array. The profiles depend on the sensor types and array geometry \cite{Zevenhoven2018}, information of which is encoded in the single-sensor MR reconstructions (Fig.~\ref{fig:profiles}). However, in conventional high-field MRI, the profiles are unstable and cannot be modeled accurately because of high-frequency and high-field effects \cite{Hoult2000a}. Furthermore, sensors used in MEG are severely incompatible with such high fields.

Conveniently for hybrid MEG--MRI, these issues are absent in ultra-low-field MRI (ULF MRI), where the magnetic fields during the measurement are on the order of 10--100 {\micro}T. At ultra-low fields, the sensitivity profiles can be modeled accurately because they are independent of the imaged sample. Consequently, the modeled profiles can be used in calibration and image reconstruction \cite{Zevenhoven2018}. ULF MRI also does not suffer from high-field-related geometrical distortions including effects of radio or microwave frequencies, tissue susceptibility or chemical shifts, and it has been demonstrated to allow imaging in the presence of metals \cite{Clarke2007}. While the resolution is still lower than in high field, detailed information from high-field images can be accurately registered even with low-resolution data \cite{guidotti2018}. Moreover, an ULF MRI scanner can be designed to have an open geometry and to operate without any sound.

For the calibration, we assume that the sensor array geometry is itself calibrated accurately \cite{Chella2012}, which is also a prerequisite for spatially accurate MEG. After this calibration, the array geometry is defined in a frame of reference which we will call the \textit{array frame}, and which also corresponds to the MEG coordinate system. However, the positions of the ULF-MRI volume elements, voxels, are conventionally determined only by the spatial encoding magnetic fields in the imaging sequence. Consequently, the \textit{image frame}, typically a grid of voxels, is independent of the array frame. To accurately position the grid of voxels in the array frame, we need another calibration, which we call ULF-MRI calibration. As the sensor array and MRI coils have fixed positions, the calibration has to be done only once for a specific imaging sequence, assuming the electronics are also stable.

The ULF-MRI calibration could be carried out, e.g., by first locating a set of fiducial points in both the array and image frame and then point-wise registering the fiducials. This approach has been demonstrated in conjunction with interleaved MEG and ULF-MRI measurements in Ref.~\cite{Magnelind2011}, with errors reported to be a few millimeters. Our method, however, is fundamentally different, 
since it utilizes a full ULF MRI signal model, including both gradient encoding and the sensitivity profiles. The method is aimed to be automatic and to enable image reconstruction into the array frame with sub-voxel and sub-millimeter spatial accuracy. 

\section{Methods}
\noindent

In the following, we introduce a model to describe voxel values in single-coil MR images from an ULF-MRI array
\cite{Zevenhoven2018}. Later in this section, we apply the model to calibrate the mapping between the MRI voxel coordinates $\boldsymbol q$ and points $\vec r$ with coordinates $\boldsymbol r  = [x, y, z]^\mathrm{T}$ in the array frame\footnote{Coordinate vectors (3$\times$1 matrices) are in boldface to distinguish them from the more abstract Euclidean vectors in three-dimensional space, denoted by arrow symbols. In this notation, arrow vectors have $\cdot$ and $\times$ products, whereas boldfaced vectors are manipulated by linear algebraic matrix operations. Arrow vectors with unit length are denoted using hat symbols.}. Finally, we describe the numerical simulations used to assess the accuracy of the calibration method.

\subsection{Signal Model}
\noindent
We use ULF-MRI-tailored Superconducting QUantum Interference Devices (SQUIDs) \cite{Luomahaara2011} to record both MEG and MRI. As the SQUID loops themselves are made small ($< 1$ mm), they are coupled with the magnetic sources via larger superconducting pickup coils. The SQUID signal is thus proportional to the magnetic flux through its pickup coil. The rectangular magnetometer pickup coils used in the simulations of this work are shown in Fig.~\ref{fig:profiles}. The pickup coils are only around 2 cm in diameter, focusing the majority of the sensitivity in a relatively small volume. More details of the working principles, array design as well as modeling the signal can be found in \cite{Zevenhoven2018}.

For a time-varying nuclear magnetization $\vec M$ in the imaged sample, the resulting flux through a pickup coil (from now on, indexed with $j$) of the sensor array can be modeled as
\begin{equation}
\Phi_j(t) = \int \vec{\mathcal{B}}_j(\vec{r}\,)\cdot\vec M(\vec r\,,t)\,d^3\vec r \,,
\label{eq:flux}
\end{equation}
where $\vec{\mathcal{B}}_j(\vec r\,)$ is the \textit{sensor field}, or the magnetization lead field of the pickup coil, and $\vec M(\vec r\,,t)$ the macroscopic magnetization at position $\vec r$ at time $t$. The volume integral is taken over the imaged object. In ULF MRI, the magnetization is increased to a measurable level by applying an additional prepolarizing pulse before the imaging sequence \cite{Clarke2007}. The magnitude of the polarization field $\vec B_\mathrm{p}(\vec{r}\,)$ determines the initial magnetization profile, $M(\vec{r}\,) \propto B_\mathrm{p}(\vec{r}\,)$, and depending on the coil design, it can be rather inhomogeneous.

In the quasi-static limit, the sensor field $\vec{\mathcal{B}}_j(\vec{r}\,)$ can be expressed as a Biot--Savart integral
\begin{equation}
\vec{\mathcal{B}}_j(\vec{r}\,) = \frac{\mu_0}{4\pi} \oint_{\partial S_j^{\,\prime}}\frac{d\vec{l}^{\,\prime} \times (\vec{r}-\vec{r}^{\,\prime})}{|\vec{r}-\vec{r}^{\,\prime}|^3} \,,
\label{eq:sensfield}
\end{equation}
where $\partial S_j^{\,\prime}$ is a path along the pickup coil wire. The fact that $\vec{\mathcal{B}}_j(\vec{r}\,)$ is the same as the magnetic field produced by a unit current in the pickup coil is an expression of the principle of reciprocity. Equation \eqref{eq:sensfield} is valid for ULF MRI, but at high fields and frequencies it becomes much more complicated as the electromagnetic properties of the object cause an altered field distribution inside the object \cite{Hoult2000a}. 

During signal acquisition, according to the Bloch equation, $\vec{M}(\vec{r},t)$ precesses around the direction of the applied magnetic field $\vec{B}(\vec{r},t)$ at the Larmor frequency $\omega(\vec{r},t) = \gamma |\vec{B}(\vec{r},t)|$, creating an oscillating flux signal in the pickup coil at the same frequency. Here, $\gamma$ is the gyromagnetic ratio. Identifying the plane perpendicular to the applied main field $\vec B_0 = |\vec B_0| \hat{e}_0$ as the complex plane, we can describe the precessing motion using a complex representation of the transverse magnetization
\begin{align}
\widetilde{M}_{\perp}(\vec{r},t) 
&= M_{\perp}(\vec{r}\,)\, \exp\left[-i\int_{0}^t \omega(\vec{r},t') dt' - i\,\phi_0(\vec{r}\,)\,\right],
\end{align}
where  $M_{\perp}(\vec{r}\,) = \sqrt{\big\vert\vec{M}(\vec{r},t)\big\vert^2 - \big[\vec{M}(\vec{r},t)\cdot \hat{e}_0\big]^2}$ is the magnitude and $\phi_0(\vec{r}\,)$ is the initial phase of the magnetization, which depends on the choice of the real axis on the precession plane. Similarly to the magnetization, we can define a complex \textit{sensitivity profile} $\beta_j(\vec{r}\,)$ \cite{Zevenhoven2018} with magnitude
$\sqrt{\big\vert{\vec{\mathcal{B}}_j(\vec{r}\,)\big\vert^2 - \big[\vec{\mathcal{B}}_j(\vec{r}\,)\cdot \hat{e}_0}\big]^2}\,$. The phase of $\beta_j(\vec{r}\,)$  depends on the direction of $\vec{\mathcal{B}}_j(\vec{r}\,)$ in the precession plane.
Using these complex quantities, the flux signal in Eq.~\eqref{eq:flux} (above zero frequency) can be written as
\begin{equation}
\Phi_j(t) = \mathrm{Re} \int \beta_j^*(\vec{r}\,) \widetilde{M}_\perp(\vec r\,,t)\,d^3\vec r \,.
\label{eq:flux2}
\end{equation}

Spatial information is encoded in the magnetization phase, which depends on the Larmor frequency as $\omega(\vec{r},t) = \omega_0 + \upDelta \omega(\vec{r},t)$, where $\omega_0 = \gamma |\vec{B}_0|$ and $\upDelta \omega(\vec{r},t)$ corresponds to the fields generated by a set of gradient coils. Demodulating the acquired flux signal $\Phi_j(t)$ by multiplying with $e^{-i\omega_0 t}$ and applying a lowpass filter (LPF), we obtain a new complex-valued signal
\begin{align}
\Psi_j(t) &= \operatorname{LPF}\left\{\Phi_j(t) e^{-i\omega_0 t}\right\} \nonumber \\ 
&= \int \beta_{j}^*(\vec r\,)M_\perp(\vec{r}\,)e^{-i\phi_0(\vec r\,)}e^{-i\phi_{\mathrm{enc}}(\vec r\,,t)}\,d^3\vec{r} \,,
\label{eq:signalintegral}
\end{align}
where $\phi_{\mathrm{enc}}(\vec{r},t) = \int_{0}^t \upDelta\omega(\vec{r},t')\, dt'$ is the phase due to spatial-encoding gradients.

The image reconstruction is performed from a finite set of data points $\Psi_{j, m}$ corresponding to the accrued phase $\phi_{\mathrm{enc}}(\vec{r}\,,t_m)$ at acquisition times $t_m$, where the index $m$ covers all the acquisitions. Normally, the image is reconstructed as numerical values assigned to image voxels, whose positions in the three-dimensional voxel array can be identified with a triple-index $\boldsymbol q_n \in \mathbb{Z}^3$. Furthermore, any point between the voxel positions can be represented by a coordinate vector $\boldsymbol q \in \mathbb{R}^3$ taking values between different $\boldsymbol q_n$. On the other hand, positions $\vec r$ in space can be determined using array-frame coordinates used for representing the geometry of the sensor array. Assuming the position vector $\vec r$ has the same origin as our array coordinate system, these coordinates are projections of $\vec r$ to the array coordinate axes $\boldsymbol r = [\vec r \,\cdot\, \hat{e}_x, \vec r \,\cdot\, \hat{e}_y, \vec r \,\cdot\, \hat{e}_z ]^\mathrm{T} = [x, y, z]^\mathrm{T}$. In addition, we define a one-to-one mapping $f$ such that
\begin{equation}
\boldsymbol r = f(\boldsymbol q) \, .
\end{equation}
The mapping $f$ depends on how the array frame is located and oriented with respect to the spatial encoding fields.

For now, we assume direct Cartesian Fourier imaging, in which the phase $\phi_{\mathrm{enc}}$ is encoded across the acquisitions using uniform gradients of $\vec{B}(\vec{r},t)\cdot \hat{e}_0$ so that the inverse discrete Fourier transform (IDFT) can be used to reconstruct the image. This means that the encoded phase can be written as $\phi_\mathrm{enc}(\boldsymbol r\, , t_m) = 2\pi \boldsymbol k_m^\mathrm{T}\boldsymbol q(\boldsymbol r)$, where $\boldsymbol q = f^{-1}(\boldsymbol r)$ and $\boldsymbol k_m$ corresponds to a normalized spatial frequency in a three-dimensional discrete Fourier transform (DFT). 

Changing the notation to the coordinate representation defined above, the data points obtained from Eq.~\eqref{eq:signalintegral} are now given by 
\begin{align}
\Psi_{j,m} =&  \int \beta_j^*(\boldsymbol r)M_\perp(\boldsymbol r) e^{ -i \phi_0(\boldsymbol r) }e^{-i2\pi \boldsymbol k_m^\mathrm{T}\boldsymbol q(\boldsymbol r)} \, d^3\boldsymbol r  \nonumber \\
=&  \int \beta_j^*(f(\boldsymbol q))M_\perp(f(\boldsymbol q)) e^{ -i \phi_0(f(\boldsymbol q)) } |\det \boldsymbol{J}(\boldsymbol q)|\; \nonumber \\ &\quad \times e^{-i2\pi \boldsymbol k_m^\mathrm{T}\boldsymbol q} \, d^3\boldsymbol q \, 
\label{eq:kdata}
\end{align}
where we have changed variables and denote the Jacobian of $f(\boldsymbol q)$ by $\boldsymbol J(\boldsymbol q)$. Here, $\Psi_{j, m}$ corresponds to a value sampled from the Fourier transform of a sensitivity-weighted image 
\begin{equation}
W_j(\mathbold q) = \beta_j^*(f(\boldsymbol q))M_\perp(f(\boldsymbol q)) e^{ -i \phi_0(f(\boldsymbol q)) } |\det \boldsymbol{J}(\boldsymbol q)|.
\label{eq:swimage}
\end{equation}

Reconstructing the data in Eq.~\eqref{eq:kdata} with IDFT, we get the value of a voxel (from now on, indexed with $n$) centered at $\boldsymbol q_n$:
\begin{align}
U_{j,n} &= \sum_m e^{i 2\pi \boldsymbol k_m^\mathrm{T}\boldsymbol q_n} \Psi_{j,m} \nonumber \\
&= \int W_j(\boldsymbol q)\sum_{m} e^{i2\pi \boldsymbol k_m^\mathrm{T}(\boldsymbol q_n-\boldsymbol q)} \,d^3\boldsymbol q \nonumber \\
&= \int W_j(\boldsymbol q)\mathrm{SRF}(\boldsymbol q_n-\boldsymbol q) \,d^3 \boldsymbol q\,,
\label{eq:psfint1}
\end{align}
where we have changed the order of summation and integration and defined the voxel function or the \textit{spatial response function} \cite{Pruessmann2006}
\begin{equation}
\mathrm{SRF}(\boldsymbol q) = \sum_{m} e^{i2\pi\boldsymbol k_m^\mathrm{T}\boldsymbol q} \,.
\end{equation}
Here, the summation is over the spatial frequencies of the three-dimensional DFT.
As can be shown, the SRF in this case is simply the periodic sinc function, also known as the Dirichlet kernel \cite{Rudin}.

\subsection{Calibration}
\label{sec:calibration}
\noindent
In the limit of low precession frequencies, the signal model can be made very accurate and stable so that a correctly reconstructed image contains no geometric distortions. However, to accurately know the origin, orientation and scaling of the image with respect to the array frame, we need calibration. Here, we present a calibration method that is based on the consistency between imaged data and the signal model in Eqs.~(\ref{eq:swimage}--\ref{eq:psfint1}).

\begin{figure}[tb]
    \centering
    \includegraphics[trim = 40mm 20mm 35mm 20mm, clip, width=0.48\textwidth]{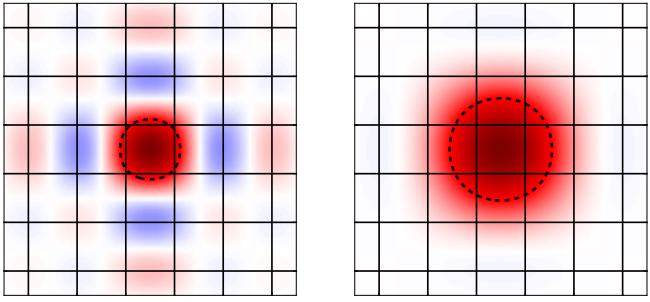}
        \caption{2D representations of (left) the Dirichlet kernel and (right) the SRF when using the Hann window, illustrating the difference in the shapes of the SRFs. Red represents positive values and blue negative ones. The black lines indicate boundaries of the cubic voxels (pixels) corresponding to the sampling frequency. The dashed lines correspond to the half-maximum values.}
        \label{fig:psf}
\end{figure}

The calibration task is to determine the mapping $f$ such that the signal equations for $U_{j,n}$ hold. For this, we use images of a phantom with ideally uniform magnetization strength $M_\perp$. However, as the prepolarization field strength $B_\mathrm{p}$ mentioned above is not uniform, this ideal situation cannot be achieved in practice. We will study the effect of inhomogeneous magnetization as a separate case.

If the energy of the SRF were concentrated close to the voxel position $\boldsymbol q_n$, it would be straightforward to make accurate approximations of the integral in Eq.~\eqref{eq:psfint1}. However, this is not the case with the Dirichlet kernel as can be seen in Fig.~\ref{fig:psf}. To enable the approximations described below, we modify the SRF by applying a window function to the $\boldsymbol k$-space data. Using a Hann window $\Pi_\mathrm{Hann}(\boldsymbol k)$, the modified SRF becomes 
\begin{equation}
\widetilde{\mathrm{SRF}}(\boldsymbol q) =  \sum_{m} \Pi_\mathrm{Hann}(\boldsymbol k_m) e^{i2\pi\boldsymbol k_m^\mathrm{T}\boldsymbol q} \,.
\end{equation}
Fig.~\ref{fig:psf} illustrates how this operation attenuates the side lobes of the SRF, while only minimally widening the main lobe.

After this modification, the phase $\phi_0(f(\boldsymbol q))$ and the Jacobian $\boldsymbol J(\boldsymbol q)$ can be approximated as being uniform within the main lobe of the SRF, which is now the effective voxel volume. Considering only the voxels whose main lobes are fully included in the phantom (interior voxels), also $M_\perp(f(\boldsymbol q))$ can be approximated uniform within each of these voxels. This allows the following approximation for the interior voxels
\begin{align}
U_{j,n} \approx\,  &  M_\perp(f(\boldsymbol q_n)) e^{i\phi_{0,n}}|\det\boldsymbol J(\boldsymbol q_n)| \nonumber \\  &\times\, \int \widetilde{\mathrm{SRF}}(\boldsymbol q_n-\boldsymbol q)   \beta_j^*(f(\boldsymbol q))\,d^3\boldsymbol q,
\label{eq:voxel_value_approx}
\end{align}
where $\phi_{0,n} = -\phi_0(f(\boldsymbol q_n))$ is the initial voxel phase.

Equation \eqref{eq:voxel_value_approx} suggests that we should convolve the sensitivities $\beta_j$ with the SRF to be able to accurately model $U_{j,n}$. As the sensitivities are spatially more rapidly varying, it is not obvious that they can be approximated uniform inside the voxel volume. However, it can be shown from Maxwell's equations that the real and imaginary components of $\beta_j$ are harmonic, i.e., $\nabla^2 \beta_j = 0$. From the mean-value property of harmonic functions \cite{Axler}, we find that convolving such a function with a spherically symmetric kernel does not affect the function values. As the main lobe of $\widetilde{\mathrm{SRF}}(\boldsymbol q)$ approximately has this symmetry, the convolution of $\beta_j$ can be evaluated as simply as $\beta_j(f(\boldsymbol q_n))$, and the approximation for the voxel value becomes
\begin{align}
U_{j,n} &\approx W_j(\boldsymbol q_n) \nonumber \\
&= \beta_j^*(f(\boldsymbol q_n)) M_\perp(f(\boldsymbol q_n)) e^{i\phi_{0,n}}|\det \boldsymbol J(\boldsymbol q_n)|.
\label{eq:voxel_value_approx2}
\end{align}

To conclude, the interior voxel values $U_{j,n}$ correspond to the profiles $\beta_j^*(f(\boldsymbol q _n))$ apart from the factor $M_\perp(f(\boldsymbol q_n))(f(\boldsymbol q)) e^{i\phi_{0,n}}|\det \boldsymbol{J}(\boldsymbol q_n)|$ and measurement noise. Note that the Jacobian is uniform unless the mapping $f$ is nonlinear. As we use only the interior voxels, the shape of the phantom does not play a role in the voxel signal model used for the calibration. In the simulations, we will use a spherical phantom, but in a real scenario, a phantom that covers the whole imaged volume would be a preferable choice.

To be able to search for the mapping that ensures the consistency of the signal model, we parametrize  $f$ with a certain set of parameters $\boldsymbol p \in  \mathbb{R}^{N_\mathrm{p}}$, the choice of which is discussed below. The problem can now be formulated as an optimization task. To this end, we select a subset of $N_\mathrm{v}$ voxels inside the phantom and denote their indices by $\nu_1, \ldots, \nu_{N_\mathrm{v}}$. For each voxel, we form a voxel vector $\boldsymbol{u}_m \in \mathbb{C}^{N_\mathrm{c}}$, where $[\boldsymbol{u}_m]_j = U_{j,\nu_m}$ and $N_\mathrm{c}$ is the number of pickup coils. In other words, each vector $\boldsymbol{u}_m$ consists of the values of the $\nu_m^{\mathrm{th}}$ voxel in $N_\mathrm{c}$ single-coil images. By concatenating vectors $\boldsymbol{u}_m$, we can represent the selection of MR data as a single vector in $\mathbb{C}^{N_\mathrm{v}N_\mathrm{c}}$
\begin{equation}
\boldsymbol{u} = \left[\boldsymbol{u}_1^\mathrm{T} \cdots \boldsymbol{u}_{N_\mathrm{v}}^\mathrm{T}\right]^\mathrm{T}\,.
\end{equation}
Similarly to the vector $\boldsymbol u$, we form a sensitivity vector
\begin{align}
\boldsymbol{s}(\boldsymbol p) &= \left[\boldsymbol{s}_1^\mathrm{T} \cdots \boldsymbol{s}_{N_\mathrm{v}}^\mathrm{T}\right]^\mathrm{T},
\end{align}
where the $j^\mathrm{th}$ element of the $m^\mathrm{th}$ subvector is $[\boldsymbol{s}_m]_j = \beta_j^*(f(\boldsymbol q_{\nu_m} |\,\boldsymbol p))\,|\det \boldsymbol{J}(\boldsymbol q_{\nu_m} |\, \boldsymbol p)|$. 

Consistency of the spatial information in the sensitivity vector $\boldsymbol s$ with the imaged data $\boldsymbol u$ is found by solving the optimization problem
\begin{equation}
\hat{\boldsymbol p} = \underset{\boldsymbol{p}}{\operatorname{argmax}}\; g(\boldsymbol{p})\,,
\label{eq:argmax}
\end{equation}
where an $g$ is an \textit{objective function}, which should be insensitive to both $M_\perp$ and the voxel phase $\phi_{0,n}$, and whose optimum should not be biased by noise in $\boldsymbol{u}$. Naturally, the choice of objective function is critical. In particular, it is important to utilize the spatial information in the phase of the sensitivity profiles while taking into account the fact that unknown factors may affect the voxel phase at the time of spatial calibration. Our previous calibration methods based on absolute-value images were able to reduce, but not fully remove the systematic bias resulting from the fact that taking the absolute value makes the noise in low-intensity voxels non-Gaussian \cite{DenDekker2014}. While emphasizing the high-signal voxels in the objective function did significantly reduce the bias, the methods were suboptimal also in terms of random error (defined in Sec.~\ref{sec:error}). 

In Appendix, we show that the phase information is exploited and the above criteria for the objective function are satisfied by
\begin{equation}
g(\boldsymbol p) = \dfrac{\sum_{n=1}^{N_\mathrm{v}} |\boldsymbol{s}_n (\boldsymbol p )^\mathrm{H} \boldsymbol{u}_n|}{\|\boldsymbol s (\boldsymbol p)\|\|\boldsymbol u \|} \, ,
\label{eq:objective_func}
\end{equation}
where $\|\cdot\|$ denotes the Euclidean vector norm and $(\cdot)^\mathrm{H}$ the conjugate transpose. 

In practice, we employ the Broyden--Fletcher--Goldfarb--Shanno (BFGS) quasi-Newton algorithm from Python's SciPy Stack for solving the nonlinear optimization problem in Eq.~\eqref{eq:argmax}. To speed up the optimization, we have implemented a semi-analytical gradient function for $g(\mathbf{p})$ and developed an interpolation scheme to avoid recalculating the profiles from scratch. Furthermore, as adjacent voxels are significantly correlated, we use only every other voxel in each dimension inside the phantom, decreasing the computation time by a factor of eight.

\begin{figure*}[tb]
    \centering
    \includegraphics[width=0.75\textwidth]{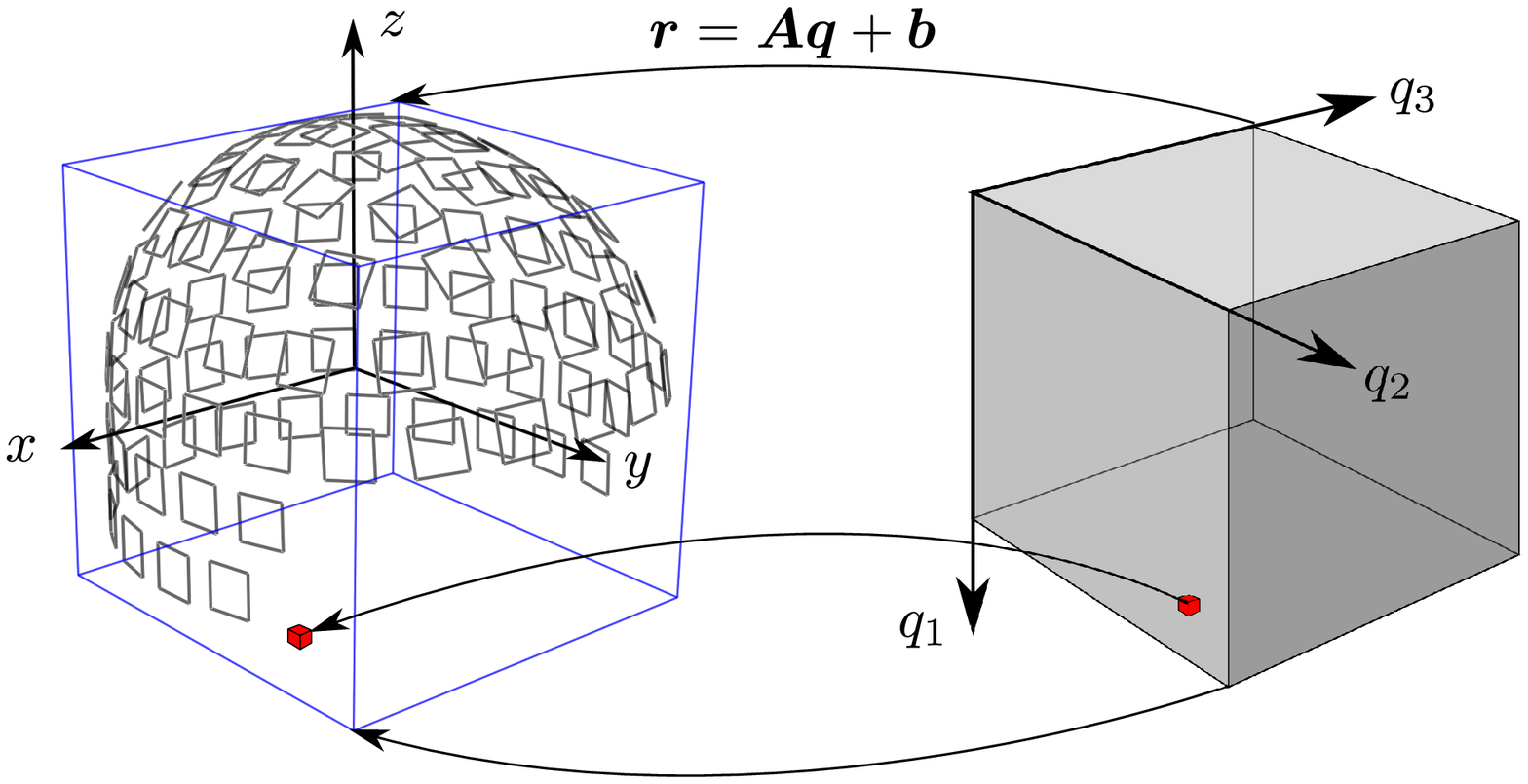}
        \caption{Schematic illustration of an affine mapping from the image coordinates $\boldsymbol q$ to array-frame coordinates $\boldsymbol r$. The image field of view on the right (gray box) maps to the volume indicated by the blue outline on the left. An example of a voxel is represented by the red cube. The axes indicate coordinate directions in each coordinate frame. The gray squares on the left represent the pickup coils of the sensor array.}
        \label{fig:mapping}
\end{figure*}

\subsection{Mapping}
\noindent 
For an accurately reconstructed ULF MRI, the mapping from voxel coordinates to Cartesian array-frame coordinates (Fig.~\ref{fig:mapping}) is given by an affine mapping
\begin{equation}
\boldsymbol r =   f_\mathrm{A}(\boldsymbol q\,|\,\boldsymbol A, \boldsymbol b) =\boldsymbol A\boldsymbol q + \boldsymbol{b} \,,
\label{eq:mappinga}
\end{equation}
where the mapping is parametrized with $\boldsymbol p \in \mathbb{R}^{12}$ containing the elements of $\boldsymbol A \in \mathbb{R}^{3\times3}$ and $\boldsymbol b\in \mathbb{R}^3$. The ideal mapping consists of translations, rotations, and scalings, which constitute nine degrees of freedom. The remaining three degrees of freedom account for shear, which should be zero unless the imaging gradients are non-orthogonal in terms of partial derivatives. In case there was any kind of affine distortion in the image, e.g., due to imperfections in the imaging sequence, this mapping would also adapt to that.

In incomplete image reconstructions, the images can suffer from distortions due to concomitant gradient fields or other field imperfections \cite{Volegov2005, Hsu2014a}. Nevertheless, when the field imperfections are small compared to $B_0$, one can model the distortions to second order with
\begin{equation}
\boldsymbol q  = h(\boldsymbol{\tilde{q}}) = \boldsymbol{\tilde{q}} + \sum_{k=1}^3 (\boldsymbol{\tilde{q}}-\boldsymbol{\tilde{q}}_0)^\mathrm{T}\boldsymbol H_k (\boldsymbol{\tilde{q}}-\boldsymbol{\tilde{q}}_0)\, \boldsymbol e_k \,,
\label{eq:mappingq}
\end{equation}
where the coordinates with tilde correspond to the undistorted coordinate system, $\boldsymbol {\tilde q}_0$ denotes an origin where there is no distortion, $\boldsymbol{H}_k$ are symmetric coefficient matrices, and $\boldsymbol e_k$ represents a unit vector along the $k^\mathrm{th}$ voxel coordinate axis. In Ref.~\cite{Kraus}, it is shown that this formulation can account for small geometric distortions due to concomitant gradients in conventional Fourier imaging at high relative gradient strengths. In this case, the mapping from the MRI to the array-frame coordinates is 
\begin{equation}
\boldsymbol{r} = f_\mathrm{A}(h^{-1}(\boldsymbol q \,|\,\boldsymbol H)\,|\,\boldsymbol A,\boldsymbol b) \,.
\label{eq:mapping2}
\end{equation}
Here, we have the inverse of the quadratic distortion because the direction of the mapping is now from the distorted coordinate system to the undistorted. For this reason, the combined mapping is not purely quadratic.

However, a pure second-order correction is found using a general quadratic mapping
\begin{equation}
\boldsymbol{r} = f_\mathrm{Q}(\boldsymbol{q} \,|\,\boldsymbol A,\boldsymbol b,\boldsymbol H) = \sum_{k=1}^3 \boldsymbol{q}^\mathrm{T}\boldsymbol H_k \boldsymbol{q}\, \boldsymbol e_k + \boldsymbol A\boldsymbol q + \boldsymbol{b} \,,
\label{eq:mapping_general}
\end{equation}
which can represent a second-order polynomial expansion of any coordinate transformation in $\mathbb{R}^3$. The parameters $\boldsymbol p \in \mathbb R^{30}$ include now also the 18 independent quadratic coefficients in  $\boldsymbol H \in \mathbb{R}^{3 \times 3 \times 3}$. After detecting (quadratic) distortions, one may want to identify and eliminate the cause of the distortion. Alternatively, the quadratic distortion can be removed by using a warped image accordingly.

\begin{figure}[tb]
    \centering
    \includegraphics[width=0.48\textwidth]{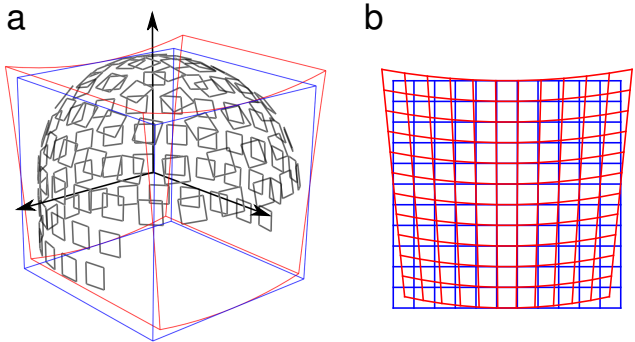}
        \caption{(a) Array-frame coordinate axes and fields of view of the MR images mapped to the array frame. The blue frame corresponds to the affine mapping and the red to the distorted mapping. The gray squares depict the magnetometer pickup coils. (b) Coordinate grid with the second-order distortion (red) and undistorted grid (blue) for comparison. The grid spacing is 16 mm, i.e., four times the voxel diameter.}
        \label{fig:fov}
\end{figure}
We emphasize that for an accurately reconstructed ULF MRI, the mapping is always affine. The additional fitting of the quadratic distortion can therefore be used to evaluate the quality of the reconstruction or to correct errors in a naively reconstructed image where effects of concomitant gradients are present.

\subsection{Non-idealities and additional considerations}
\noindent
The calibration technique was designed to be independent of external parameters such as the reconstruction technique, and the required data should only be the channel-wise reconstructed MR images. In an actual imaging sequence, a possible non-ideality affecting the image quality can then also degrade the calibration,  unless taken into account. However, as the overall aim is to maximize the image quality while solving also these issues, we have left modeling of their effects out of this work. For example, field distortions related to pulsing MRI coils can be reduced by coil design and Dynamical Coupling for Additional dimeNsions (DynaCAN) \cite{Zevenhoven2013degauss, Zevenhoven2015}, and the MRI electronics should have high precision in order not to affect the image quality \cite{Zevenhoven2014amp}.

Even if unknown, we have conveniently taken the possibly non-uniform phase into account in the design of the objective function. However, there are at least two complicating aspects  modeled already in Eq.~\eqref{eq:voxel_value_approx2} but ignored in the construction of the objective function in the Appendix. These are (a) the inhomogeneity in the magnetization of the phantom due to inhomogeneous polarization field and (b) the uncertainty in the direction of the $\vec{B}_0$ field with respect to the array frame. 


To address (a), we can take an iterative approach. In the first calibration, we assumed a homogeneous polarization, and due to this approximation, some error may remain in the calibration parameters $\boldsymbol p = \boldsymbol p_1$. However, reliable estimates for the sensitivity vectors $\boldsymbol s_n(\boldsymbol p_1)$ are now available and the underlying magnetization profile can be estimated voxel-wise as \cite{Zevenhoven2018}
\begin{equation}
M_n(\boldsymbol p_1) = \dfrac{|\boldsymbol{s}_n(\boldsymbol p_1)^\mathrm{H}\boldsymbol u_n|}{\|\boldsymbol{s}_n(\boldsymbol p_1)\|^2}\,.
\end{equation}
Next, update the sensitivity model to $M_n(\boldsymbol p_1) \boldsymbol s_n(\boldsymbol p)$, i.e., include the estimated magnetization inhomogeneity in the sensitivity profiles. Then, iterate the calibration by updating the magnetization estimate $M_n(\boldsymbol p_i)\to M_n(\boldsymbol{p}_{i+1})$ until converged. Results of this procedure are presented in the next section.

The other unknown factor (b) is the direction of $\vec{B}_0$. The profiles $\beta_j(\boldsymbol r)$ calculated in the \textit{array frame} depend on the direction of $\vec{B}_0$, but as $\vec{B}_0$ has no reference in this frame, its direction should also be calibrated together with the mapping parameters. For the sake of computation time and unambiguity of the results, we assumed this direction to be known in the noise simulations and tested optimizing it in separate cases.

\subsection{Error Analysis}
\label{sec:error}
\noindent
To study the effects of image noise, approximations in the modeling, and the overall robustness of the calibration method, we used a series of numerical simulations. We generated 3D MR images for the 102 magnetometer pickups in a helmet-shaped array based on a standard MEG configuration from Elekta Oy (Helsinki, Finland). An affine mapping [Eq.~\eqref{eq:mappinga}] was fixed between the voxel coordinates and the array frame so that the field of view (FOV) of the MR image matched the extent of the sensor array. To assess the detection of geometric distortion, we constructed another mapping with a quadratic geometric distortion [Eq.~\eqref{eq:mapping2}] corresponding to a gradient-field variation of $1/5\,B_0$ over half of the FOV, which roughly matches  the maximum relative gradient strengths used in Ref.~\cite{Vesanen2013}. Fields of view corresponding to both mappings as well as a depiction of the second-order distortion are shown in Fig.~\ref{fig:fov}.

The MR images were simulated using Eq.~\eqref{eq:kdata}, where the sensitivity profiles were calculated based on the Biot--Savart integral, using the exact analytical formula from the Appendix of \cite{Zevenhoven2018} and a given direction of $\vec{B}_0$. $M_\perp$ was uniform inside a sphere with radius 85 mm, which fits well inside the sensor helmet, leaving a 25-mm distance to the closest sensor positions. The phase $\phi_0(\boldsymbol q)$ was chosen to be uniform in the simulation, although in a realistic scenario, there may be slight deviations from that. The image was sampled from a field of view of 192$\times$192$\times$192 mm$^3$ and the voxel size was 4$\times$4$\times$4 mm$^3$. The continuous SRF convolution in Eq.~\eqref{eq:psfint1} was calculated as a multiplication in the $\boldsymbol k$ space. We first mimicked the continuous Fourier transform in Eq.~\eqref{eq:kdata} by oversampling the FOV by a factor of eight and using a special memory-efficient version of the Fast Fourier Transform (FFT) \cite{Bailey1991} to calculate only the required samples in the center of the $\boldsymbol k$ space. We then added white Gaussian noise to the samples, after which we applied the 3D Hann window to the data.

Effects of noise were studied by simply running the calibration algorithm several times, each time with a different realization of the noise. Statistics were calculated over 50 calibration runs for different signal-to-noise ratios (SNR) defined as
\begin{equation}
\mathrm{SNR} = \sqrt{\dfrac{\| \boldsymbol{u} \|^2}{N_\mathrm{v}N_\mathrm{c}\sigma^2}} \,,
\end{equation}
where $\boldsymbol{u}$ is a noiseless image vector consisting of voxels inside the phantom, $N_\mathrm{v}$ is the number of voxels, $N_\mathrm{c}$ is the number of pickup coils (i.e., $N_\mathrm{v}N_\mathrm{c}$ is the dimension of $\boldsymbol{u}$), and $\sigma^2$ is the variance of the noise in each voxel.

We analyzed the quality of the calibration by determining the systematic calibration error (SCE) and the random calibration error (RCE), which can be thought of as a 3D generalization of the mean and standard deviation of the errors, respectively, as explained in the following. The positional error vector in the $k^\mathrm{th}$ calibration run was defined as
\begin{equation}
\boldsymbol{d}_k(\boldsymbol{r}) = \boldsymbol{r} - f_k(f^{-1}(\boldsymbol{r}))\, ,
\end{equation}
where $f$ is the true mapping and $f_k$ the mapping obtained from the $k^\mathrm{th}$ calibration run. Using this definition, the systematic calibration error was estimated as
\begin{equation}
\mathrm{SCE}(\boldsymbol{r}) = \left\|\dfrac{1}{K}\sum_{k=1}^K\boldsymbol{d}_k\right\| = \left\|\overline{\boldsymbol{d}}\right\|
\end{equation}
and the random calibration error as
\begin{equation}
\mathrm{RCE}(\boldsymbol{r}) =  \sqrt{ \dfrac{1}{K} \sum_{k=1}^K  \left(\boldsymbol{d}_k - \overline{\boldsymbol{d}}\,\right)^\mathrm{T} \left(\boldsymbol{d}_k - \overline{\boldsymbol{d}}\,\right) }\, .
\end{equation}
Here, the overbar denotes the arithmetic mean. 

In addition to the simulations described above, we ran some tests to study the effect of inhomogeneous magnetization. We added a magnetization profile proportional to a polarization field strength $B_\mathrm{p}(\boldsymbol r)$ fitted to corresponding fluxgate magnetometer measurements inside the sensor helmet. The fitted field strength had an inhomogeneity $(B_{\mathrm{p},\mathrm{max}}-B_{\mathrm{p},\mathrm{min}})/B_{\mathrm{p},\mathrm{max}}$ of around 60\% over the spherical phantom.

\section{Results}
\noindent
We examined the calibration results for SNR values ranging from 0.5 to infinity (zero noise). In all cases, the optimization algorithm converged in the vicinity of the correct optimum. One indication of the robustness of the method is that we employed no prior information of the image scale or orientation, but used zero as the initial guess for each parameter. This corresponds to a transformation that maps all voxels to the origin of the array frame, which is in the middle of the sensor array. 

\begin{figure}[bt]
	\centering
    \includegraphics[width=0.5\textwidth]{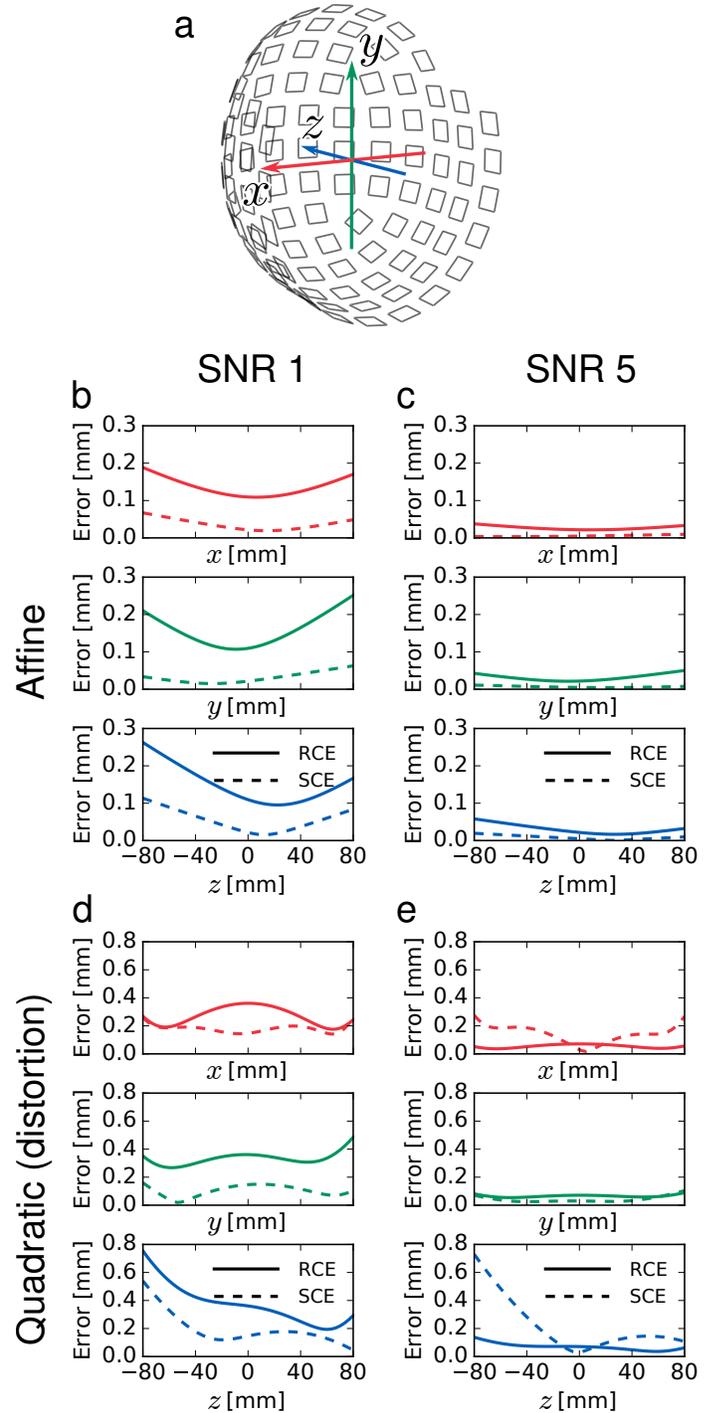}
        \caption{Calibration errors plotted  inside the sensor helmet on the three Cartesian coordinate axes depicted in (a): $x$ (red), $y$ (green), and $z$ (blue). (b) Affine calibration, $\mathrm{SNR}$ = 1. (c) Affine calibration, $\mathrm{SNR}$ = 5. (d) Quadratic calibration, $\mathrm{SNR}$ = 1. (e) Quadratic calibration, $\mathrm{SNR}$ = 5.}
\label{fig:errors_axes}
\end{figure}

In this section, we show results for two very low SNR values $\mathrm{SNR} = 1$ and $\mathrm{SNR} = 5$, because these values were already sufficient for high spatial accuracy. Calibrations were obtained for both (a) well-reconstructed undistorted images simulated using an affine mapping [Eq.~\eqref{eq:mappinga}] and (b) distorted images simulated using the mapping in Eq.~\eqref{eq:mapping2}.   Fig.~\ref{fig:errors_axes} shows the calibration error statistics plotted along the axes shown in Fig.~\ref{fig:errors_axes}a.

In the case of correctly reconstructed, undistorted images, only the twelve affine parameters were fitted. For $\mathrm{SNR} = 1$ (Fig.~\ref{fig:errors_axes}b) $\mathrm{RCE} <$ 0.3~mm and $\mathrm{SCE} <$ 0.2~mm are much smaller than the voxel diameter of 4 mm. In Fig.~\ref{fig:errors_axes}c, we see that, with the SNR increasing from 1 to 5, RCE decreases roughly by the same factor. Such correlation suggests that measurement noise in the estimate is dominating the actual systematic error. We can only conclude that SCE is always much smaller than RCE.

\begin{figure*}[tb]
    \centering
    \includegraphics[width=\textwidth]{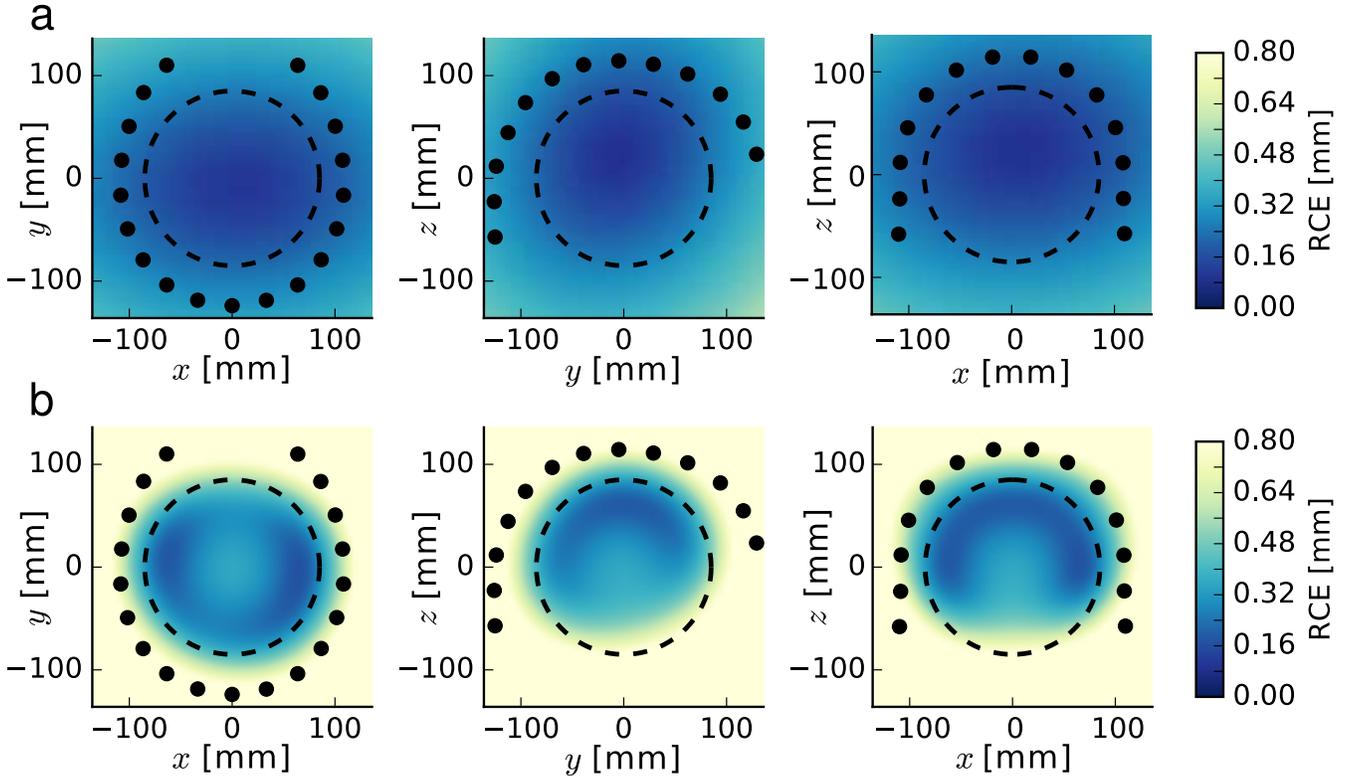}
        \caption{Random calibration error of voxel positions, plotted on the planes spanned by the axes in Fig.~\ref{fig:errors_axes}e (a)   Affine calibration with $\mathrm{SNR}=1$ (b) Quadratic calibration with $\mathrm{SNR}=1$. The dots represent the closest sensor positions projected on the plane and the dashed line denotes the boundary of the spherical phantom. Note that the colormap saturates at 0.8 mm.}
        \label{fig:erplanes}
\end{figure*}

For detecting distortions in incorrectly reconstructed images, we fitted the quadratic mapping [Eq.~\eqref{eq:mapping_general}] with 18 additional coefficients. The error statistics are shown in Figs.~\ref{fig:errors_axes}d and e. Comparing the calibration errors to those in pure affine calibration in Figs.~\ref{fig:errors_axes}b and c, we see that the calibration accuracy of the quadratic mapping is more prone to noise. Furthermore, we observe increased systematic error, especially on the negative $z$ axis where the sensor array is least sensitive. This is due to the fact that, for detecting the distortions, we use the purely quadratic mapping [Eq.~\eqref{eq:mapping_general}], which cannot completely represent the inverse-quadratic distortion [Eq.~\eqref{eq:mapping2}] in the simulated images. Nevertheless, since the calibration errors are mostly below 0.4 mm (up to 0.8 mm far from the sensors), the distortions are well detected even in high-noise conditions.

For a more qualitative analysis, in Fig.~\ref{fig:erplanes}, we have plotted the estimated random calibration error for $\mathrm{SNR}$ = 1 on $xy$, $yx$ and $xz$ planes. In the case of affine calibration (Fig.~\ref{fig:erplanes}a), the smallest random calibration error can be found near the sensors inside the spherical phantom. The effect is even more pronounced in Fig.~\ref{fig:erplanes}b, which corresponds to the case of quadratic mapping. In conclusion, independent of the mapping, using our calibration method, the smallest calibration error is found where the sensor array is most sensitive. 

\begin{figure}[tb]
\centering
    \includegraphics[width=0.48\textwidth]{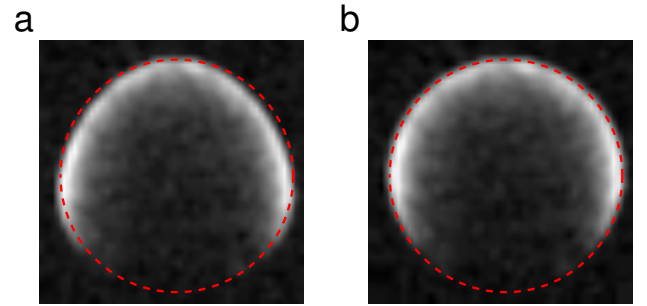}
        \caption{Correction of minor geometric distortion for simulated MR images. (a) Slice of a sum-of-squares image of the spherical phantom with distortion due to incorrect reconstruction and (b) the same slice warped using quadratic parameters found from a calibration with $\mathrm{SNR} = 1$. The red dashed circle is a visual aid to show the true shape of the phantom.}
        \label{fig:warp}
\end{figure}

As mentioned previously, after detecting a distortion in the reconstructed image, it can be corrected by warping it back to the original geometry. Fig.~\ref{fig:warp} demonstrates the second-order geometry correction of a distorted image using a mapping according to Eq.~\eqref{eq:mapping_general}, which was determined using images with $\mathrm{SNR} = 1$. In the distorted image, the deviation from the true geometry was around two voxels at the edges of the sphere. Still, the calibration error was much smaller than the voxel size. Using the calibrated mapping, the image could thus be warped back quite well to the true spherical geometry.

\begin{figure}[tb]
    \centering
    \includegraphics[width=0.48\textwidth]{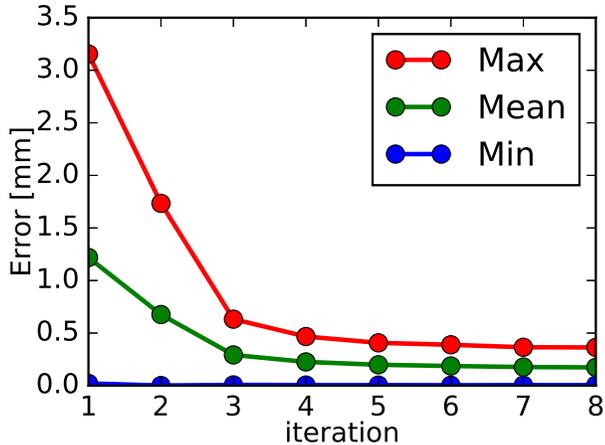}
        \caption{Maximum, mean, and minimum calibration errors over the calibration phantom for a set of simulated images with additional inhomogeneity and SNR=1. After three iterations the maximum error has decreased below one millimeter.}
        \label{fig:iteration}
\end{figure}

In addition to noise simulations, we ran separate calibrations for images with additional inhomogeneity, modeling the effect of nonuniform prepolarization, as described in the previous section. In the uncorrected case, the calibration error increased to around 3.5 mm, independent of noise. As described previously, this effect can be alleviated by estimating the inhomogeneity from the images and adding it to the sensitivity model. In Fig.~\ref{fig:iteration}, we see the effect of this procedure in the calibration error as a function of iterations, for phantom images with $\mathrm{SNR}=1$. After a few iterations, sub-millimeter maximum error inside the calibration phantom was again achieved.

To test another nonideality, we ran a test case with spatially rapidly varying phase $\phi_0(\boldsymbol q)$ added to the simulated images. These images contained quadratically increasing phase towards the edges of the FOV with maximum phase difference of $2\pi$. In this case, we noted an increase in the calibration error but it was limited to 0.5 mm within the FOV. 

Finally, we let the direction of the precession plane (determined by $\vec{B}_0$) be optimized in conjunction with the mapping parameters, utilizing the same objective function. We ran several test cases with incorrect initializations of this direction. First, keeping the incorrect direction fixed while optimizing the mapping led to erroneous calibrations. However, when also the direction parameters were optimized, the direction robustly turned towards the true direction of $\vec{B}_0$. We did not see notable differences in the calibration error compared to the ideal case, although the computation time was longer.

With the present implementation, calibrating the affine mapping takes about one minute on a desktop computer (Intel Xeon quad-core CPU at 3.2 GHz with 8 GB of memory). Adding the quadratic distortion parameters increased the computation time of the optimization to around 3 minutes. 

\section{Discussion}
\noindent
Assuming an affine mapping between the MRI and array-frame coordinates, we performed the calibration of ULF MRI using simulated images by maximizing the given objective function with respect to the twelve affine parameters. When images of all the 102 magnetometers in the sensor array are in use, this is a massively overdetermined problem, which results in negligible spatial error even when low-SNR images are used in the calibration. The vast amount of data can be used to fit even more degrees of freedom, e.g., to detect distortions in the images. Fitting 18 additional quadratic parameters to distorted images was also shown to work, although the calibration error far from the sensors diverged faster. In principle, any kind of deformation model with a sufficiently small number of degrees of freedom could be applied for detecting nonlinear geometric distortions.

In addition to the random error, the calibration always contains systematic error, which can only be removed by perfectly modeling the acquired signals and designing an objective function free of noise bias. The objective function was shown to perform well also in terms of systematic error even when significant amount of noise ($\mathrm{SNR}=1$) was added to the images. Although the sensitivity model itself \cite{Zevenhoven2018} is very accurate in ULF MRI, additional inhomogeneity due to the initial prepolarization can slightly alter the voxel values and cause error up to a few millimeters in the calibration. However, by taking the inhomogeneity into account in the profiles and iterating the method for a few times, we were able to eliminate this error.

Errors in the sensor array geometry could also play a role in ULF MRI calibration, but this was not studied as a calibrated sensor geometry is also a prerequisite for accurate MEG. In Ref.~\cite{Dabek2012b}, it is shown how, after calibrating ULF MRI to the array frame, phantom data could be used to calibrate the positions and orientations of the pickup coils. However, any error in the ULF MRI calibration could then produce errors in the sensor-array geometry. To keep the calibration of the array geometry independent of the ULF MRI calibration, we assumed the sensor array was calibrated, e.g., with a geometrically accurate electrical phantom ~\cite{Chella2012,oyama2015,ilmoniemi2009}. Nonetheless, it is not known if alternating iterations of spatial ULF MRI calibrations and sensor array calibrations using only ULF MRI data could be used to perform both calibrations.

Also, timing errors in the MRI sequence and gradient non-linearities beyond the concomitant components were not studied in this work, as they can be reduced by careful design and implementation of the measurement system. The effect of system imperfections left unmodeled should be confirmed in an actual experiment in which the calibration method can be tested against some ground truth. This is left for future work with a next-generation MEG--MRI device, which we are currently developing. However, it should be noted that any affine or quadratic distortion originating from such imperfections can be compensated for using the mapping to the calibrated distortion-free array frame.

As shown in the previous section, high calibration accuracy is only guaranteed near the sensitive volume of the array. This is especially true for a sensor array that does not cover the whole head. Based on simulations, for a flat array, the objective function may also have another optimal location for the image on the other side of the array, so care must be taken when calibrating ULF MRI with such an array. On the other hand, distributing fewer sensors around the head can increase the random error compared to the case of the full head array, but the error is globally more restricted than with a flat array.

To make the computations tractable, we assumed that the concomitant gradients have been compensated for to yield no distortion or that they generate only geometric distortions. As shown in Ref.~\cite{Volegov2005}, increasing relative gradient strengths adds blurring at the edges of the FOV due to inconsistencies in phase encoding. In addition, the assumption of second-order geometric distortion breaks down. In such a case, accurate reconstruction can only be performed using a specialized method that takes into account the known concomitant gradients \cite{Hsu2014a, Nieminen2010}. Mere geometric distortions can be determined using this calibration method with a suitable non-linear mapping.

Regarding the image simulations, we took the continuous nature of the MR measurement into account by heavily oversampling the field of view, which requires a significant amount of memory and computation time. The voxel size of $4\times4\times4\, \mathrm{mm}^3$ was chosen partly because a similar sizes were used in the first brain images acquired by the system \cite{Vesanen2013}, but the main reason was that smaller voxels would have required much more computational resources. In Ref.~\cite{Guerquin-Kern2012a}, uniform-phantom computations were made much more efficiently by approximating the sensitivity profiles inside an ellipse with a set of basis functions and analytically calculating the Fourier integrals for each of the functions. This approach may also be adapted for calculating the $\boldsymbol k$-space signal from our profiles inside the spherical phantom so that the computational burden could be reduced. 

Alone, calibrating ULF-MR images to the array frame with high accuracy guarantees a perfect co\hyp{}registration only if the head stays still during the measurements. This could be achieved by fixing the head position with subject-specific head-casts \cite{Troebinger2014a}, although they are not easily applicable in daily MEG workflow. Otherwise, the head movement must be tracked and compensated for in the data. In a typical MEG setup, the head tracking is implemented with a set of localization coils attached to the scalp \cite{Ahlfors1989, Uutela2001, DeMunck2001}. In contrast to conventional MEG, after the ULF-MRI coordinate calibration, we do not need the absolute positions of the localization coils with respect to the head or each other, but only the change of the coil positions with respect to the measurement device. This further eliminates the errors related to 3-D digitizer operation \cite{Engels2010}. An alternative way to track the head would be to use the fact that the MRI sensitivity profiles encode the head displacement differently in each measurement channel when recording NMR signals from the head. With a sufficient model of the underlying magnetization distribution of the head, the movement parameters could possibly be inferred from, e.g., a free induction decay (FID) signal. Although movement tracking and compensation are necessary, these methods are left outside the scope of this work.

\section{Conclusions}
\noindent
In this work, an accurate method was designed for spatially calibrating ULF-MRI data using a hybrid MEG--MRI system and the consistency between the MRI signal model and calibration-phantom data. This eliminates the conventional MEG--MRI co-registration step. The method finds affine parameters for mapping the voxel indices to coordinates in the sensor-array frame and can be used to fit an additional quadratic mapping to detect and correct for minor geometric distortions in incorrectly reconstructed images. After this, the array frame can be used as a common (laboratory) frame to represent both MEG and MRI data. The method was shown to work robustly and accurately using simulated MRI data. Sub-voxel and sub-millimeter calibration accuracy was achieved even in very low SNR conditions. Our approach eliminates all sources of the conventional co-registration error and can reduce overall errors in spatial alignment to a negligible level.


\appendix[Derivation of the objective function]
\label{sec:app1}
\noindent
As explained in Sec.~\ref{sec:calibration}, to solve the calibration problem, the spatial information in the image vector $\boldsymbol u$ should be made consistent with the sensitivity vector $\boldsymbol{s} = \boldsymbol{s}(\boldsymbol p)$. If the only unknown in the calibration were the white Gaussian noise in the image, this condition could be straightforwardly estimated by minimizing $\|\boldsymbol u-\boldsymbol s(\boldsymbol p)\|^2$. Unfortunately, there are also scale and phase ambiguities between the vector elements [see Eq.~\eqref{eq:voxel_value_approx}]. 

The scale ambiguity is assumed to be uniform and can be eliminated by simply normalizing $\boldsymbol u$ and $\boldsymbol s$. The ambiguity in phase is trickier since signal cancellation can occur if the phase is not uniform across the voxels. One solution could be taking absolute values of the vector elements, but this would lead to Rician noise \cite{DenDekker2014}, which has a biased mean in low-SNR voxels and can thus shift the optimal $\boldsymbol p$ at high noise levels. However, close to the optimum, the phase of the voxel can be estimated using an expression for the array-reconstructed $n^{\mathrm{th}}$ complex voxel value \cite{Zevenhoven2018}, which for uniform-variance noise uncorrelated across the sensors reduces to
\begin{equation}
v_n = \dfrac{\boldsymbol s_n^\mathrm{H}\boldsymbol u_n}{\mathbf{s}_n^\mathrm{H}\boldsymbol s_n}\,,
\end{equation}
where $\boldsymbol s_n$ and $\boldsymbol u_n$ are subvectors of $\boldsymbol s$ and $\boldsymbol u$ corresponding to the $n^\mathrm{th}$ voxel. The phase factor can then be estimated as
\begin{equation}
e^{i\hat{\phi}_{0,n}} = \dfrac{v_n}{|v_n|} 
= \dfrac{\boldsymbol s_n^\mathrm{H}\boldsymbol u_n}{|\boldsymbol s_n^\mathrm{H}\boldsymbol u_n|}\, .
\label{eq:phase_estimate}
\end{equation}

Defining the phase-corrected image vector as $\boldsymbol u'$ where $\boldsymbol u'_n = \boldsymbol u_n e^{-i\hat{\phi}_{0,n}}$, we can perform the calibration by minimizing the squared error
\begin{align}
\left\|\dfrac{\mathbf{u'}}{\|\mathbf{u'}\|} - \dfrac{\mathbf{s}}{\|\mathbf{s}\|}\right\|^2 &= \dfrac{\|\mathbf{u'}\|^2}{\|\mathbf{u'}\|^2} - \dfrac{\mathbf{s}^\mathrm{H}\mathbf{u'}}{\|\mathbf{s}\|\|\mathbf{u'}\|} -  \dfrac{\mathbf{u'}^\mathrm{H}\mathbf{s}}{\|\mathbf{u'}\|\|\mathbf{s}\|} + \dfrac{\|\mathbf{s}\|^2}{\|\mathbf{s}\|^2} \nonumber \\
&= 2\left(1 -\mathrm{Re}\dfrac{\mathbf{s}^\mathrm{H}\mathbf{u'}}{\|\mathbf{s}\|\|\mathbf{u'}\|}\right),
\label{eq:squared_difference}
\end{align}
where we note that maximizing the complex inner product $\mathrm{Re}\dfrac{\boldsymbol{s}^\mathrm{H}\boldsymbol{u'}}{\|\boldsymbol{s}\|\|\boldsymbol{u'}\|}$ yields an optimum at the same point. By rewriting the inner product using the voxel vectors $\boldsymbol u'_n$ and $\boldsymbol s_n$ and inserting the phase estimates, we get
\begin{align}
\mathrm{Re}\dfrac{\boldsymbol{s}^\mathrm{H}\boldsymbol{u'}}{\|\boldsymbol{s}\|\|\boldsymbol{u'}\|} &= 
\mathrm{Re}\dfrac{\sum_{n=1}^{N_\mathrm{v}}\boldsymbol{s}_n^\mathrm{H}\boldsymbol{u}'_n}{\|\boldsymbol{s}\|\|\boldsymbol{u'}\|} \nonumber \\
&= \dfrac{1}{\|\boldsymbol{s}\|\|\boldsymbol u\|}\mathrm{Re}\sum_{n=1}^{N_\mathrm{v}} \boldsymbol{s}_n^\mathrm{H}\boldsymbol u_n e^{-i\hat{\phi}_{0,n}} \nonumber \\
&= \dfrac{1}{\|\boldsymbol{s}\|\|\boldsymbol u\|} \mathrm{Re}\sum_{n=1}^{N_\mathrm{v}}\dfrac{ |\boldsymbol{s}_n^\mathrm{H}\boldsymbol u_n|^2}{|\boldsymbol{s}_n^\mathrm{H}\boldsymbol u_n|}  \nonumber \\ 
&= \dfrac{\sum_{n=1}^{N_\mathrm{v}} |\boldsymbol{s}_n^\mathrm{H}\boldsymbol u_n|}{\|\boldsymbol{s}\|\|\boldsymbol u\|}\, ,
\end{align}
where the last expression gives the objective function in Eq.~\eqref{eq:objective_func}.

Although the phase estimate in Eq.~\eqref{eq:phase_estimate} is correct only in the vicinity of the optimum, where the sensitivity vector matches the true sensitivities, we have noted in simulations that it does not affect the convergence of the method. In the end, the objective function is very similar to the cosine measure of absolute value vectors, but exploits also the fact that, at the optimum, the phases in $\boldsymbol{s}_n$ should match $\boldsymbol u_n$ apart from the local magnetization phase $\phi_{0,n}$.

\bibliographystyle{IEEEtran}
\bibliography{library_final.bib}   

\end{document}